\documentclass{appolb}
\usepackage{amsmath}
\usepackage{amssymb,amsthm}
\usepackage{graphicx}
\begin{document}

\author{\L{}ukasz Skowronek${}^{\mathrm{a}}$\thanks{e-mail:cwirus@poczta.onet.pl}
and
P.\ F.\ G\'ora${}^{\mathrm{a,b}}$\thanks{e-mail:gora@if.uj.edu.pl}
\address{${}^{\mathrm{a}}$Marian Smoluchowski Institute of Physics,
Jagellonian University, Reymonta~4, 30-059 Krak\'ow, Poland\\
${}^{\mathrm{b}}$Mark Kac Complex Systems Research Center,
Jagellonian University, Reymonta~4, 30-059 Krak\'ow, Poland}}

\title{Chaos in Newtonian iterations: Searching for zeros which are not there}

\date{February 28, 2007}

\maketitle

\begin{abstract}
We show analytically that Newtonian iterations, when applied to a~polynomial
equation, 
have a positive topological entropy. In a specific example of an attempt to 
``find'' the real solutions of the equation $x^2+1=0$, we show that the Newton method
is chaotic. We analytically find the invariant density and show how this problem relates
to that of a piecewise linear map.
\end{abstract}


\section{Introduction}

Suppose we want to numerically solve a nonlinear equation $f(z) = 0$, where $z\in\mathbb{R}$
or $z\in\mathbb{C}$. If only calculating 
the derivative of $f$ is possible, perhaps the famous Newton (or Newton-Raphson)
method \cite{numrec} is the first method that comes to mind. Starting from some
$z_0$, this method uses the following iterations:

\begin{equation}\label{start}
z_{n+1} = z_n - \frac{f(z_n)}{f^\prime(z_n)} = \eta(z_n)\,.
\end{equation}

\noindent If this iteration converges, it usually does so very fast. Because of that,
and because of its simplicity, the Newton method is one of the most frequently
used numerical methods ever. Yet it is widely known that even if $f(z)$ 
is a~low-order polynomial, boundaries between basins of attractions of different 
roots can be
very complicated. Indeed, many pieces of popular software use this particular
property of the Newtonian iterations to produce aesthetically pleasing fractals.

\begin{figure}
\begin{center}
\includegraphics[scale=0.95]{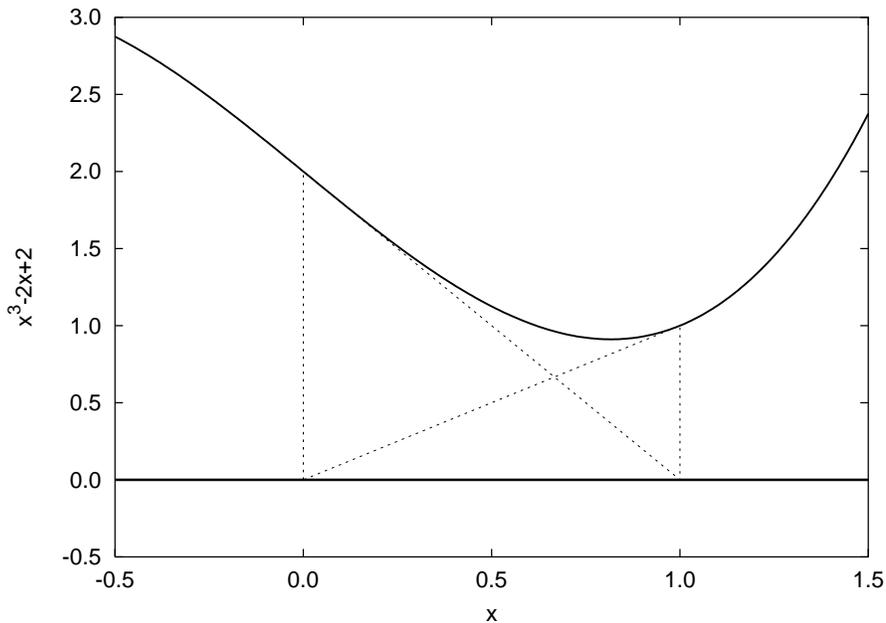}
\end{center}
\caption{A $2$-cycle generated by the Newton method applied to
the equation \hbox{$x^3-2x+2=0$}.}
\label{stabilny}
\end{figure}

The Newton method fails if it hits a~zero of the derivative,
or when the denominator in \eqref{start} vanishes, unless the zero of the
derivative corresponds to a multiple root. It is less commonly known
that the Newton method also fails to find a root if it forms a~multicycle.
Many textbooks that teach the Newton method do not even mention this fact and
those that do make the impression that multicycles in the Newton method are
a~rare and unimportant peculiarity. In reality, however, constructing examples that display
multicycles is quite easy. A~reader may verify, for instance, that the points
$\{0,1\}$ form a~stable $2$-cycle, shown in Fig.~\ref{stabilny}, 
in the Newton method applied to the polynomial
$x^3-2x+2$. Several questions then
arise: How many such multicycles are there? Is it possible to encounter them
in practical numerical applications of the Newton method? Is there a connection
between the multicycles and, say, basins of attraction?

We may regard consecutive iterations of the Newton method as a dynamical system.
Multicycles generated by the Newtonian iterations are the periodic orbits
of this dynamics. Since the seminal works of Cvitanovic and co-workers 
\cite{Cvitanovic}, it is known that the unstable periodic orbits (UPOs) carry
important information about the dynamical system that generates them. Yet to
extract this information, one needs to know at least how many periodic orbits
there are.

We shall discuss these questions in the case of $f(z)$ being a polynomial.

This paper is organized as follows: In Section~\ref{sec2} we show that the
Newtonian iterations applied to a polynomial equation can generate infinitely
many multicycles and we calculate the topological entropy associated with them.
In Section~\ref{x2} we discuss a particularly simple case of the polynomial
$z^2+1$. We show that dynamics resulting from the Newton method, when restricted
to the real axis, is chaotic and equivalent to the dynamics generated by 
a~piecewise linear map. We explicitly calculate the invariant density
of the Newtonian dynamics in this case. Since the number of chaotic systems
whose invariant densities are known analytically is fairly limited, this
result is very interesting, at least from a~pedagogical point of view.
In Section~\ref{x2m} we partially generalize these results to the case
of polynomials $z^{2m}+1$, $m\geqslant2$. Conclusions are given in 
Section~\ref{concl}. The Appendix contains a proof of the theorem that we use
to calculate the topological entropy.

\section{Polynomials and multicycles}\label{sec2}

Let $P(z)$ be a complex polynomial of order $n$:

\begin{equation}\label{newton:P}
P(z) = \sum\limits_{s=0}^n a_s z^s\,.
\end{equation}

\noindent Suppose the polynomial \eqref{newton:P} has $n_d$ distinct 
roots\footnote{Distinct roots of a polynomial must not be confused with roots of muliplicity 
one. For example, the polynomial $x^2(x-1)^2(x-2)$, of order five, has \textit{three} 
distinct roots.}. We solve the polynomial equation

\begin{equation}\label{newton:rownanie}
P(z)=0
\end{equation}

\noindent numerically by Newtonian iterations. This procedure defines a function
\begin{equation}\label{newton:etageneral}
\eta(z) = z - \frac{P(z)}{P^\prime(z)}\,.
\end{equation}

\noindent
Consecutive iterates of~\eqref{newton:etageneral} satisfy

\begin{equation}\label{newton:xk}
\eta^k(z) = \eta^{k-1}(z)-\frac{P\left(\eta^{k-1}(z)\right)}{P^\prime\left(\eta^{k-1}(z)\right)}\,,\quad
k=1,2,\dots
\end{equation}

\noindent
with $\eta^0(z)\equiv z$.
The following Theorem gives a more detailed characteristics of the iterates of the
function~\eqref{newton:etageneral}. It is pivotal in our subsequent discussion:

\newtheorem{twierdzenie}{Theorem}

\begin{twierdzenie}\label{twierdzenie1}
$\forall\; k\geqslant1$
\begin{equation}\label{newton:twierdzenie}
\eta^k(z)=z - \frac{P(z)}{P^\prime(z)}\,\frac{A_k(z)}{B_k(z)}
\end{equation}
\end{twierdzenie}
\noindent where $A_k$, $B_k$ are polynomials of order $n_d^k-n_d$, where $n_d$ is the
number of \textit{distinct} roots of the polynomial $P(z)$. 

An elementary, but rather lenghty and technical, proof of this Theorem is given in the Appendix.

Theorem~\ref{twierdzenie1} is very useful when we consider multicycles
generated by the function~\eqref{newton:etageneral}. Specifically, all points belonging to 
a~$k$-cycle satisfy

\begin{equation}\label{newton:kcycle1}
\eta^k(z)=z\,.
\end{equation}

\noindent Using Eq.~\eqref{newton:twierdzenie} we get that either $P(z)=0$,
which means that $z$ is a fixed point of the function~\eqref{newton:etageneral},
or that

\begin{equation}\label{newton:Ak}
A_k(z) = 0\,.
\end{equation}

\noindent This equation has $n_d^k-n_d$ solutions on the complex plane. If $k$ is prime,
then there are $N_k=(n_d^k-n_d)/k$ different $k$-cycles. If $k$ is not prime, we need
to subtract ``spurious'' $k$-cycles; for example, two rounds over a 2-cycle can be
mistakenly taken for a~4-cycle. Eventually, for the number of $k$-cycles that are generated 
by the Newton method applied to a polynomial with $n_d$ distinct zeros, we obtain

\begin{equation}\label{Nk}
N_k = \frac{1}{k}\left(n_d^k - n_d - \sum\limits_{j\in D_k} j\cdot N_j\right),
\end{equation}

\noindent where $D_k$ is a set of all proper divisors of $k$ (if $k$ is prime, 
$D_k=\emptyset$). These multicycles are periodic orbits of the dynamics generated by
successive iterations of the function \eqref{newton:etageneral}. A~limited number of the 
multicycles may be stable, but as in any dynamical system, an overwhelming majority of 
them are unstable and form the unstable periodic orbits (UPOs) of the dynamics.

The number of multicycles grows very rapidly with $k$; for example, if
$n_d=5$, there are approximately $4.5\cdot10^{10}$ $17$-cycles. To quantify this
observation, we can calculate the topological entropy \cite{entropy} of 
the dynamics generated by $\eta(x)$ by appropriately counting the UPOs \cite{Greborgi}:

\begin{equation}\label{newton:entropy}
S = \lim\limits_{k\to\infty} \frac{1}{k} \ln N_k = \ln n_d\,.
\end{equation}

\noindent $S>0$ for all $n_d\geqslant2$. It is well known that if a map acting on 
an interval has a positive topological entropy, it is chaotic \cite{Franzowa}.
In a more general setting, if the topological entropy is positive, we can expect 
\textit{some} form of chaotic behaviour to show up in the dynamics.

We can see that the Newton method, when applied to a~polynomial equation, leads
to an extremely rich structure of multicycles. But where are they? If any of them
is stable, as in our example in the Introduction, it can show up in a practical 
application of the Newtonian iterations.
The unstable majority live on fractal boundaries between basins of attraction.
It is the dynamics on these boundaries that is the chaotic behaviour whose presence
is indicated by a~positive value of the topological entropy. It is also interesting to 
see that if a polynomial has multiple roots, it has less multicycles than a~polynomial
of the same order and with all roots distinct: It is the number of \textit{distinct}
roots, rather than the degree of the polynomial or the number of all the roots,
that determines the value of the topological entropy.

\section{The equation $x^2+1=0$}\label{x2}

In general, multicycles generated by the Newton method live on complicated geometric
objects somewhere on the complex plane. There is, however, one nontrivial example
where the locale of the multicycles can be pinpointed quite accurately. Consider the
equation

\begin{equation}\label{newton:z2}
z^2+1=0\,.
\end{equation}

\noindent It generates the following Newtonian dynamics:

\begin{equation}\label{newton:etaz}
\eta(z) = \frac{1}{2}\left(z-\frac{1}{z}\right).
\end{equation}

\noindent It can be proved that if we start with any $z_0$ such that $\text{Im}(z_0)>0$,
the Newtonian iterations converge to $z_\infty=+i$. Similarly, if we start with 
$\text{Im}(z_0)<0$, the Newtonian iterations converge to $z_\infty=-i$. Therefore, all
the multicycles of the function \eqref{newton:etaz} and all the points that eventually
end up on them must lie on the real axis.

\begin{figure}
\begin{center}
\includegraphics[scale=0.95]{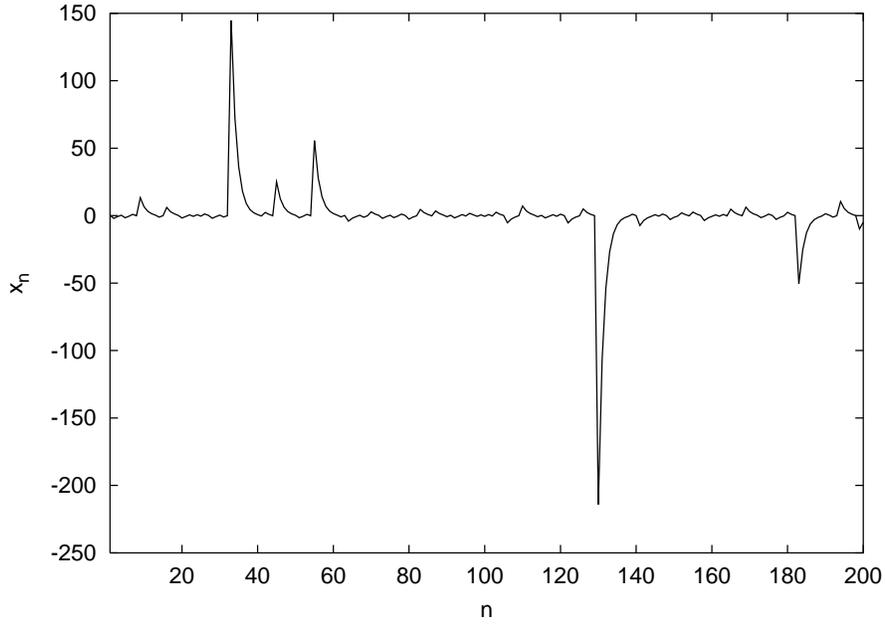}
\end{center}
\caption{A typical trajectory generated by the map \eqref{newton:eta}. The
trajectory spends most of its time in the vicinity of $x=0$, but from time to
time it makes large excursions, and then slowly relaxes towards zero. If one
waits long enough, arbitrarily large excursions can be observed.}
\label{fig-trajektoria}
\end{figure}

We shall, therefore, discuss properties of the map \eqref{newton:etaz} restricted to the 
real axis:

\begin{equation}\label{newton:eta}
\eta(x) = \frac{1}{2}\left(x-\frac{1}{x}\right),\quad x\in\mathbb{R}\,.
\end{equation}

\noindent Formally speaking, this map results from an attempt
to find the real zeros of the equation $x^2+1=0$. There are no such zeros, but
that does not necessarily mean that properties of the map \eqref{newton:eta}
are not interesting. Fig.~\ref{fig-trajektoria} shows a~typical trajectory generated
by this map.

\subsection{Properties of the map \eqref{newton:eta}}

We shall list the most important properties of the map \eqref{newton:eta}:

\begin{itemize}
\item This function is singular in $x=0$ and it increases monotonically for
both $x<0$ and $x>0$. It does not have a fixed point.
\item Each point has two preimages. The preimages of a~point $x$ satisfy
\begin{equation}\label{newton:preimages}
x^\pm = x \pm \sqrt{x^2+1}\,.
\end{equation}
\item There are countably many multicycles (UPOs). For each multicycle, there are
countably many points that eventually fall onto it.
\item $x=0$ corresponds to the escape to infinity. There are countably many
points that are eventually mapped into $x=0$ and escape to infinity.
\item The union of all the multicycles, points
that lead to them, and points that eventually escape to infinity, is also countable
and dense in $\mathbb R$. This leaves us with uncountably many points
that neither escape, nor fall on a multicycle, but form a~``truly chaotic'' orbit
of the dynamical system~\eqref{newton:eta}.

\end{itemize}

Simulations suggest that the map \eqref{newton:eta} has an invariant density
that does not depend on the starting point, provided this point does not escape
to infinity. This eliminates only countably many points, and in practice only
three of them: $x\in\{-1,0,1\}$ (the preimages of $\pm1$ are irrational and inaccessible
in numerical simulations). Indeed, to compute the invariant density, denoted
here by $\rho(x)$, we use
the Frobenius-Perron equation \cite{Beck}:

\begin{equation}\label{newton:FP}
\rho(x) = \int\limits_{-\infty}^\infty \delta\left(\eta(x)-y\right)\rho(y)\,dy\,.
\end{equation}

\noindent Using Eq.~\eqref{newton:preimages}, this equation takes the explicit form

  \begin{eqnarray}
  \sqrt{x^2+1}\,\rho(x) &=& \left(x+\sqrt{x^2+1}\right)\rho\left(x+\sqrt{x^2+1}\right) 
  \nonumber\\
                        &-& \left(x-\sqrt{x^2+1}\right)\rho\left(x-\sqrt{x^2+1}\right).
  \label{frob2}
  \end{eqnarray}

It can be verified by a direct substitution that the Lorentzian distribution

\begin{equation}\label{newton:lorentz}
\rho(x) = \frac{1}{\pi(1+x^2)}
\end{equation}

\noindent solves the Frobenius-Perron equation \eqref{frob2}. Thus the map \eqref{newton:eta}
joins the elite club of maps whose invariant densities are known explicitly. Slowly
decaying tails of the invariant density \eqref{newton:lorentz} explain the 
large deviations from zero made by the chaotic trajectory.

\subsection{The stereographic projection}

It is interesting to see that the stereographic projection convertss the function 
\eqref{newton:eta} into a piecewise linear map on the unit interval. If we substitute

\begin{equation}\label{tangens}
x = \tan\pi\phi\,,
\end{equation}

\noindent we obtain

  \begin{equation}\label{trygonometria}
  \eta(x) = \cot 2\pi\phi = \tan \pi\left(2\phi+\frac{1}{2}\right).
  \end{equation}

\noindent
Given the periodicity of the trigonometric function, we can see that the map
\eqref{newton:eta} is equivalent to 

  \begin{equation}\label{zeta}
  \zeta(\phi) = 2\phi + \frac{1}{2} \mod 1\,.
  \end{equation}

Properties of the map \eqref{zeta} are easy to find and they nicely illustrate 
the multicycles.

\begin{itemize}

\item The map \eqref{zeta} is chaotic with the Lyapunov exponent $\lambda=\ln2$.
This map has a flat invariant density, $\rho(\phi)=1$, $\phi\in[0,1]$.

\item $\phi=1/2$ is the fixed point of the map \eqref{zeta}. The fixed point
corresponds to an escape to infinity in the language of $x$ given by \eqref{tangens}.

\item If we represent any $\phi\in[0,1]$ by its binary expansion, the map \eqref{zeta}
acts on it by (i) the binary shift, (ii) rejecting the highest bit, and (iii) flipping
the remaining highest bit. For example,

\begin{equation}\label{przyklad}
0.1100101\dots_2\to
0.000101\dots_2\,,
\end{equation}

\noindent where the subscript indicates a binary expansion. Thus, all points with finite
binary expansions eventually converge to $0.1_2=1/2$, i.e., escape to infinity, points
with periodic (starting form some point) binary expansions belong to multicycles
or settle on one after a~finite number of steps, and points
with nonperiodic, infinite binary expansions lie on the chaotic trajectory. 
Thus the rational numbers either escape to infinity under the action of \eqref{zeta},
or end up on multicycles. As the transformation \eqref{tangens} maps rational
numbers from the unit interval to irrational numbers in $\mathbb{R}$, encountering
a~true multicycle of \eqref{newton:eta} during a~computer simulation is
impossible. However, the fact that computer calculations are performed with a finite
precision, leads to the conclusion that in a~computer experiment, the map
\eqref{newton:eta} behaves as if it were periodic, with a~fairly large period,
almost regardless of the starting
point. This is a~well known fact about pseudo-random sequences.

\item We can count
the $k$-cycles by counting nontrivially different binary sequences of the length $k$.
For example, there are $2^k$ possibilities of distributing $\{0,1\}$
among $k$ sites. Two constant sequences must be excluded, and as all
cyclic permutations are equivalent, we arrive, if $k$ is prime,
on $(2^k-2)/k$. (If $k$ is not prime, the result of $(2^k-2)/k$ is fractional.)
This result, and its generalizations to non-prime $k$'s, are the same
as those given by Eq.~\eqref{Nk} with $n_d=2$.

\end{itemize}

\begin{figure}
\begin{center}
\includegraphics[scale=0.65]{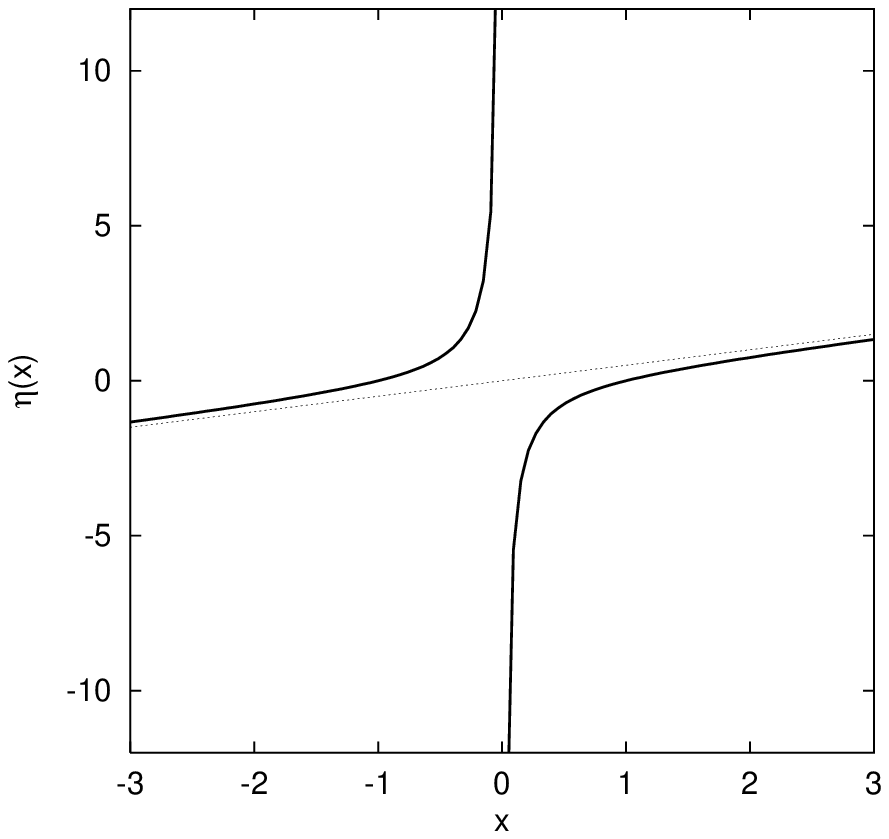}\includegraphics[scale=0.65]{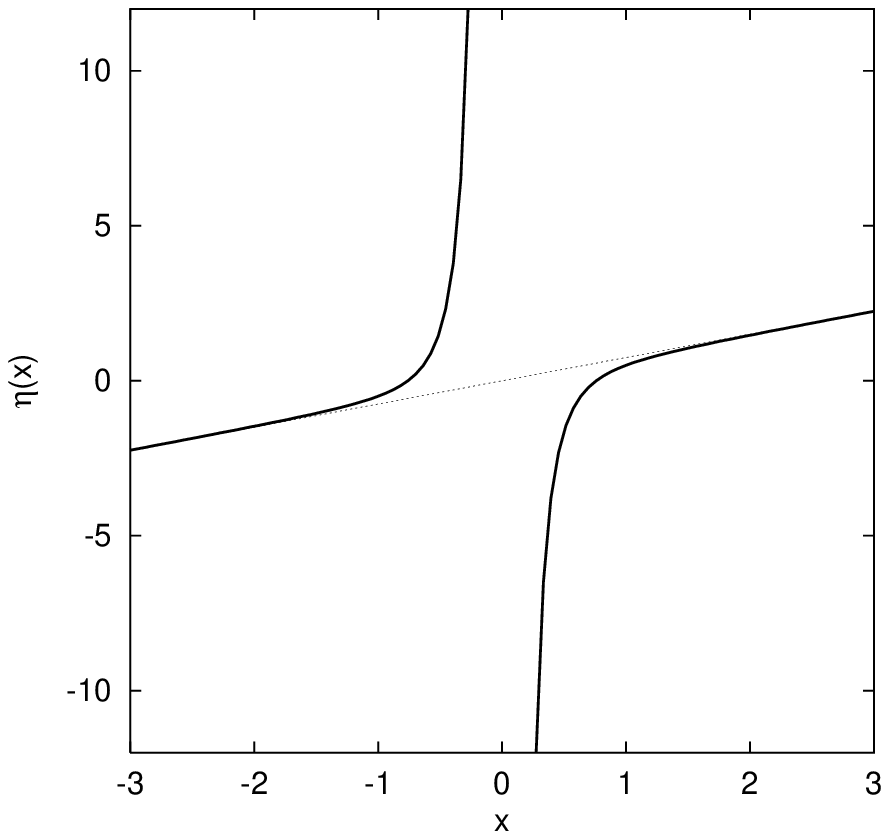}
\includegraphics[scale=0.65]{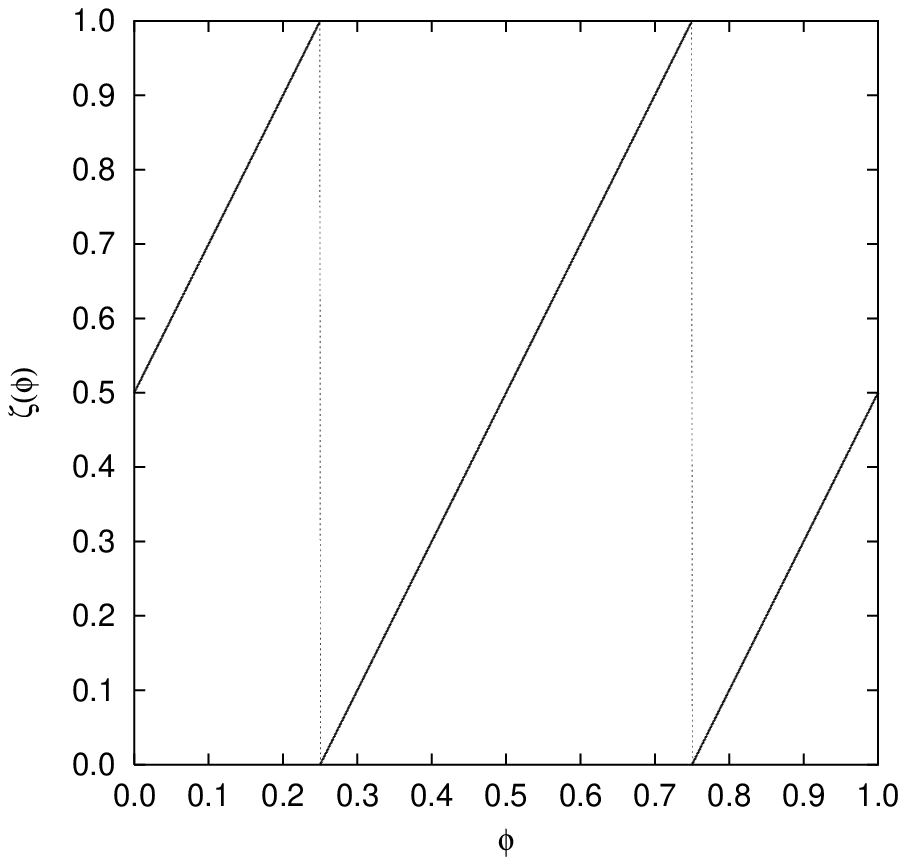}\includegraphics[scale=0.65]{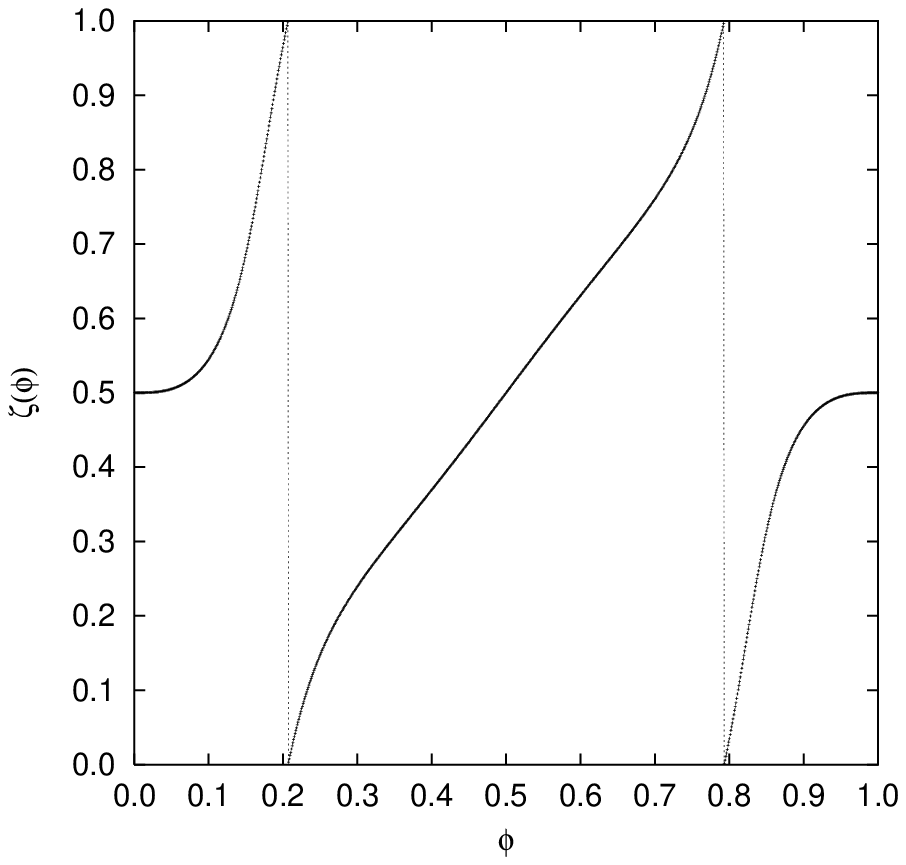}
\end{center}
\caption{The map \eqref{newton:eta} (top left) and the map \eqref{zeta} (bottom left),
compared to the map \eqref{tildeeta} with $m=2$ (top right) and the corresponding
map on the unit interval, obtained numerically (bottom right).}
\label{fig-mapy}
\end{figure}

\section{The equation $x^{2m}+1=0$}\label{x2m}

The equation

\begin{equation}\label{newton:z2m}
z^{2m}+1=0\,,\quad z\in\mathbb{C}\,,\quad m\geqslant2
\end{equation}

\noindent is a natural generalization of the equation \eqref{newton:z2}.
However, the multicycles generated by the Newtonian iterations resulting from
the equation \eqref{newton:z2m} are not restricted to the real axis. Indeed,
in case of Eq.~\eqref{newton:z2m}
the Newton method generates identical dynamics on each line $e^{i\pi l}$
on the complex plane, with
$l=0,1,\dots\,m-1$, but there are also fractal boundaries between the basins
of attraction that do not lie on these lines and, apparently, a~majority
of the multicycles are located there. If the Newtonian iterations are
restricted to the real axis, they are the iterates of the function

\begin{equation}\label{tildeeta}
\eta(x) = \left(1-\frac{1}{2m}\right)x -\frac{1}{2m\,x^{2m-1}}\,,\quad x\in\mathbb{R}\,.
\end{equation}

\noindent This function has \textit{qualitatively} the same properties
as the function \eqref{newton:eta}: it is singular at $x=0$, grows monotonically
from $-\infty$ to $\infty$ in both domains $x<0$ and $x>0$, and each point has two
preimages. Therefore, the number
of its multicycles (again, on the real axis) must be the same as that of 
\eqref{newton:eta}. If we use the substitution \eqref{tangens}, the function
\eqref{tildeeta} does not convert to a~piecewise linear map. Fig.~\ref{fig-mapy}
compares the maps \eqref{newton:eta} and \eqref{zeta} with the map
\eqref{tildeeta} with $m=2$ and the counterpart of $\zeta(\phi)$, obtained numerically
from \eqref{tildeeta} with \eqref{tangens}.

\begin{figure}
\begin{center}
\includegraphics{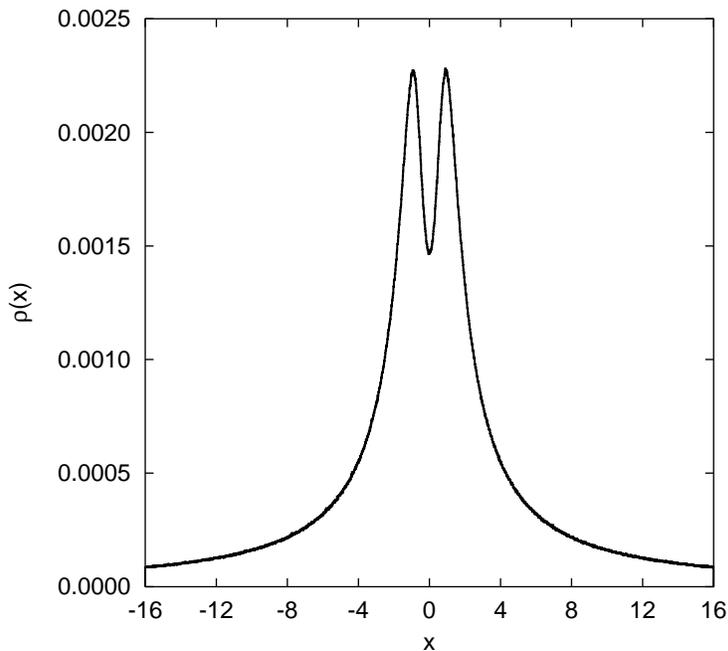}
\end{center}
\caption{The invariant density corresponding to the function \eqref{tildeeta} with $m=2$,
obtained numerically. The tails behave like $\sim|x|^{-4/3}$, in a~full agreement with
our theoretical predictions.}
\label{fig-tails}
\end{figure}

We now turn to the invariant density.
In the asymptotic regime, $x\gg1$, preimages of a point $x$ satisfy

\begin{subequations}\label{przyblizone}
\begin{eqnarray}
\eta^{-1}(x) &\simeq&x^+= \frac{2m}{2m-1}x\,,\\
\eta^{-1}(x) &\simeq&x^-= -\frac{1}{(2mx)^{1/(2m-1)}}\,.
\end{eqnarray}

\noindent Moreover, if an invariant density exists and is continuous,
it must satisfy

\begin{equation}
\rho(x^-) \simeq \rho(0)\,.
\end{equation}
\end{subequations}

\noindent Thus, if $x\gg1$, the Frobenius-Perron equation \eqref{newton:FP} takes
the approximate form

\begin{equation}\label{frobasymptotic}
x^{\frac{2m}{2m-1}}
\left[\rho(x) - \frac{2m}{2m-1}\,\rho\left(\frac{2mx}{2m-1}\right)\right]
=\frac{\rho(0)}{(2m-1)(2m)^{1/(2m-1)}} \,.
\end{equation}

\noindent Because of the symmetry of the function \eqref{tildeeta}, 
the invariant density, if it exists, must satisfy $\rho(x)=\rho(-x)$.
The right-hand side of \eqref{frobasymptotic} is constant. Thus, the
left-hand side can be constant only if

\begin{equation}\label{asymptotic}
\rho(x) \sim \text{const}\cdot |x|^{-2m/(2m-1)}\,,\quad |x|\to\infty\,.
\end{equation}

\noindent 
Fig.~\ref{fig-tails} shows the invariant density, found numerically, corresponding 
to the function \eqref{tildeeta} with $m=2$. The tails agree perfectly with
the theoretical prediction \eqref{asymptotic}.
Simulations show that for all $m\geqslant2$, the invariant density is bimodal,
and that the valley between the peaks gets deeper as $m$ increases.

\section{Conclusions}\label{concl}

We have shown that the Newton method, when applied to a polynomial equation that
has more than one distinct solution,
generates infinitely many multicycles, and we have calculated the topological
entropy associated with them.
We have also shown that the map \eqref{newton:eta} is chaotic, equivalent to
a~piecewise linear map on the unit interval, and we have explicitly
found its invariant density. This result is interesting by itself as the 
number of maps with analytically known invariant densities is very limited.
Because the invariant density in question coincides with the Lorentzian
probability distribution, one could use the function \eqref{newton:eta}
as a~basis of a pseudo-random generator of the latter.

From a practical point of view, hitting an unstable multicycle is virtually
impossible, although it is well known that starting the Newtonian iterations
in a close vicinity of a fractal boundary between the basins of attractions,
where the unstable multicycles live, can slow down the convergence significantly.
In some cases, stable multicycles can also appear and they can prevent the
Newton method from converging to a~root; the polynomial presented in 
Fig.~\ref{stabilny} provides one such example.
It would be interesting to see what are the criteria for the existence of stable
multicycles. In practice, however, if a multicycle appears to form, the 
Newtonian iterations should be interrupted by a couple of steps taken with a~different
method, and with the damped Newton method in particular. This usually breaks the
multicycle that has started to form, but can lead to new multicycles in their own
right. Discussing this point goes beyond the scope of the present paper.
We should finally mention that the Newton method is \textit{not} the best method
for a~numerical search for zeros of a~polynomial. There are other algorithms, better
tailored to polynomials. From those the Laguerre method \cite{numrec,ralston},
coupled with the deflation of the polynomial, is now regarded as the method of choice.
Despite many arguments in favour of the latter method, the Newton method is
probably most commonly used for the task, and this is why a~study of its UPOs structure
is, in our opinion, important.

To end on a lighter note, we have shown that a numerical search for zeros of a~polynomial
with the Newton method can be quite chaotic, in particular when the zeros are just not there.

\L{}S thanks Mr Wojciech Brzezicki for a helpful discussion.
This work was supported in part by the Marie Curie Actions Transfer of
Knowledge project COCOS (contract MTKD-CT-2004-517186).

\appendix

\section{}

\begin{proof}

We give a proof of Theorem~\ref{twierdzenie1} from Section~\ref{sec2} in this Appendix.
First observe that if the polynomial $P(z)$ has multiple roots, there exists a~polynomial
$Q(z)$ such that $P(z) = Q(z)R(z)$ and $P^\prime(z)=Q(z)S(z)$, where $\text{deg}\,R(z)=n_d$,
$\text{deg}\,S(z)=n_d-1$, $n_d$ is the number of \textit{distinct} roots of the polynomial
$P(z)$ and the fraction $R(z)/S(z)$ cannot be further cancelled. (If all roots of $P(z)$
are distinct, $n_d=n$ and $Q(z)\equiv1$.) Therefore (cf.~\eqref{newton:twierdzenie})

\begin{equation}\label{ulamek}
\eta^k(z) = z - \frac{R(z)}{S(z)}\,\frac{A_k(z)}{B_k(z)}\,.
\end{equation}

\noindent To prove the theorem, we need to show that
\begin{itemize}
\item[I.] The polynomials $A_k(z)$, $B_k(z)$, generated according to \eqref{newton:xk},
have degrees $n_d^k-n_d$, and
\item[II.] The polynomials $R(z)$, $S(z)$, $A_k(z)$, $B_k(z)$ do not have common roots,
from which it follows that the fraction in \eqref{ulamek} cannot be further cancelled.
\end{itemize}
Both parts of the proof proceed by induction. Note that $A_1(z)\equiv B_1(z)\equiv1$,
so the theorem holds for $k=1$. For the sake of the notation, let us assume that

\begin{equation}
R(z)=\sum\limits_{j=0}^{n_d} r_j z^j\,.
\end{equation}

\subsection*{Part I}

Suppose the theorem holds for some $k\geqslant1$ and calculate 

\begin{gather}
R\left(z^{(k)}\right) 
=
R\left(z - \frac{R(z)}{S(z)}\,\frac{A_k(z)}{B_k(z)}\right)
=
r_0 + \sum\limits_{s=1}^{n_d} r_s\left(z - \frac{R(z)}{S(z)}\,\frac{A_k(z)}{B_k(z)}\right)^s
\nonumber\\
{}=
r_0 + \sum\limits_{s=1}^{n_d} r_s \sum\limits_{l=0}^s{s\choose l}z^{s-l}(-1)^l
\left(\frac{R(z)}{S(z)}\,\frac{A_k(z)}{B_k(z)}\right)^l
\nonumber\\
{}=
\sum\limits_{s=0}^{n_d} r_s z^s + 
\sum\limits_{s=1}^{n_d} r_s \sum\limits_{l=1}^s{s\choose l}z^{s-l}(-1)^l
\left(\frac{A_k(z)}{S(z)B_k(z)}\right)^l(R(z))^l
\nonumber\\
{}=
R(z)\left[1+\sum\limits_{s=1}^{n_d} r_s \sum\limits_{l=1}^s{s\choose l}z^{s-l}(-1)^l
\left(\frac{A_k(z)}{S(z)B_k(z)}\right)^l(R(z))^{l-1}\right]
\nonumber\\
\noalign{\noindent(note that $l-1\geqslant0$)}
{}=
\frac{R(z)}{(S(z)B_k(z))^{n_d}}
\Biggl[(S(z)B_k(z))^{n_d} + {}
\nonumber\\
\label{newton:Rzk}
\sum\limits_{s=1}^{n_d} r_s (S(z)B_k(z))^{n_d-s}
\sum\limits_{l=1}^s{s\choose l}(-1)^l(zS(z)B_k(z))^{s-l}(A_k(z))^l
(R(z))^{l-1}\Biggr].
\end{gather}

\noindent Similarly
\begin{gather}
S\left(z^{(k)}\right) = 
S\left(z - \frac{R(z)}{S(z)}\,\frac{A_k(z)}{B_k(z)}\right)
\nonumber\\
\label{newton:Rprimezk}
{}=
\frac{1}{(S(z)B_k(z))^{n_d-1}}
\sum\limits_{s=0}^{n_d-1} (s+1)r_{s+1} (S B_k)^{n_d-1-s}
(z\,S B_k - RA_k)^s\,.
\end{gather}

\noindent Therefore
\begin{equation}\label{newton:RtoRrime}
\frac{R\left(z^{(k)}\right)}{S\left(z^{(k)}\right)} =
\frac{R(z)}{S(z)}\,\frac{{\cal N}(z)}{B_k(z){\cal D}(z)}\,,
\end{equation}
where ${\cal N}(z)$, ${\cal D}(z)$ are certain polynomials whose forms are implied
in Eqns.~\eqref{newton:Rzk} and~\eqref{newton:Rprimezk}. What are the orders of these
polynomials? $\text{deg}\,S=n_d-1$ and, by assumption, $\text{deg}\,A_k = 
\text{deg}\,B_k = n_d^k-n_d$. We thus have

\begin{eqnarray}
\text{deg}\,(S B_k)^{n_d-s} &=& (n_d-s)(n_d-1+n_d^k-n_d)\nonumber\\
&=&n_d^{k+1}-n_d^ks-n_d+s\,,\\
\text{deg}\,(z\,S B_k)^{s-l}(A_k)^l R^{l-1} &=&
(s{-}l)(1{+}n_d{-}1{+}n_d^k{-}n_d)+l(n_d^k{-}n_d) +(l{-}1)n_d\nonumber\\
&=&n_d^ks -n_d\,.
\end{eqnarray}
Thus the term under the sum over $s$ in~\eqref{newton:Rzk} is of the order
$n_d^{k+1}-n_d^ks-n_d+s+n_d^ks=n_d^{k+1}-2n+s$, which takes a maximal value of $n_d^{k+1}-n_d$
for $s=n_d$. Therefore
\begin{equation}\label{newton:order1}
\text{deg}\,{\cal N}(z) = n_d^{k+1}-n_d\,.
\end{equation}
Similarly, because $\text{deg}\,z\,S=\text{deg}\,R=n_d$,
\begin{eqnarray}
\text{deg}\,(S B_k)^{n_d-s-1}(z\, S B_k {-} RA_k)^s &=&
(n_d{-}s{-}1)(n_d{-}1{+}n_d^k{-}n_d)\nonumber\\
&&{}+s(n_d{+}n_d^k{-}n_d)\nonumber\\
&=&n_d^{k+1}-n_d^k-n_d+s+1\,,
\end{eqnarray}
which takes a maximal value of $n_d^{k+1}-n_d^k$ for $s=n_d-1$. Therefore
\begin{equation}\label{newton:order2}
\text{deg}\,{\cal D}(z) = n_d^{k+1}-n_d^k\,.
\end{equation}

We finally have

\begin{eqnarray}
\eta^{k+1}(z) &=& \eta^k(z) - \frac{R\left(\eta^k(z)\right)}{S\left(\eta^k(z)\right)}
= z - \frac{R(z)}{S(z)}\,\frac{A_k(z)}{B_k(z)} -
\frac{R(z)}{S(z)}\,\frac{{\cal N}(z)}{B_k(z){\cal D}(z)}
\nonumber\\\label{ulamek2}
&=& z - \frac{R(z)}{S(z)}\,
\frac{{\cal D}(z)A_k(z) + {\cal N}(z)}{B_k(z){\cal D}(z)}
= z - \frac{R(z)}{S(z)}\,\frac{A_{k+1}(z)}{B_{k+1}(z)}\,.
\end{eqnarray}
The last statement defines polynomials $A_{k+1}(z)$, $B_{k+1}(z)$.
Using Eqns.~\eqref{newton:order1} and~\eqref{newton:order2}, it is
now easy to verify that $\text{deg}\,A_{k+1}=\text{deg}\,B_{k+1}=n_d^{k+1}-n_d$,
provided the last fraction in \eqref{ulamek2} cannot be cancelled.

By direct calculations, it is easy to show that

\begin{subequations}\label{AkBk}
\begin{eqnarray}
A_{k+1}(z) &=& A_k(z)B_k^{n_d-1}(z)S^{n_d-1}(z)
\,S\left(z-\frac{R(z)}{S(z)}\,\frac{A_k(z)}{B_k(z)}\right)
\nonumber\\
\label{Ak}
&+&B_k^{n_d}(z)S^{n_d}(z)R^{n_d-1}(z)\,R\left(z-\frac{R(z)}{S(z)}\,\frac{A_k(z)}{B_k(z)}\right),
\\
\label{Bk}
B_{k+1}(z) &=& B_k^{n_d}(z)S^{n_d-1}(z)\,S\left(z-\frac{R(z)}{S(z)}\,\frac{A_k(z)}{B_k(z)}\right).
\end{eqnarray}
\end{subequations}
\noindent
Note that we have already shown that $A_{k+1}(z)$, $B_{k+1}(z)$ \textit{are} polynomials.

\subsection*{Part II}

To complete the proof, we still need to show that the polynomials $R(z)$, $S(z)$, 
$A_k(z)$, $B_k(z)$ do not have common roots. The polynomials $R(z)$, $S(z)$ do not 
have common roots by construction. Suppose the polynomials $R(z)$, $S(z)$, 
$A_k(z)$, $B_k(z)$ do not have common roots for some $k\geqslant1$.

\subsubsection*{$R(z)=0$}

Let $R(z)=0$. Then $S(z)\not=0$ and $B_k(z)\not=0$, and because of \eqref{Bk}, 
$B_{k+1}(z)=(B_k(z)S(z))^{n_d}\not=0$. Furthermore, if $R(z)=0$, then, by virtue of
\eqref{Ak}, $A_{k+1}(z)=A_k(z)(B_k(z)S(z))^{n_d-1}S(z)\not=0$, because $A_k(z)\not=0$, either.
In other words, $R(z)$ does not have common roots with the other polynomials.

\subsubsection*{$S(z)=0$}

If $S(z)=0$, then $z$ is a zero of the derivative of the original polynomial $P(z)$ that
is not a multiple root of the latter. In this case the Newton iterations diverge and
we cannot use the expressions \eqref{AkBk}. We need to use the definitions impiled in
\eqref{ulamek2} instead; they are still valid by the argument of continuity. 
We have $B_{k+1}(z)=B_k(z){\cal D}(z)$, and if $S(z)=0$,
this reduces to $B_{k+1}(z)=s_{n_d{-}1}(R(z)A_k(z))^{n_d-1}\not=0$, 
where $s_{n_d{-}1}$ is the highest-order coefficient in the polynomial $S(z)$,
cf.~Eq.~\eqref{newton:Rprimezk} above. For the remaining polynomial we have
$A_{k+1}(z)=A_k(z){\cal D}(z)+{\cal N}(z)$ which, for $S(z)=0$, reduces to
$A_{k+1}(z)=(-1)^{n_d} R^{n_d-1}(z)A_k^{n_d}(z)(r_{n_d}{-}s_{n_d{-}1})$, where $r_{n_d}$
is the highest coefficient in $R(z)$. Thus if $S(z)=0$, $A_{k+1}(z)$ could vanish
only if $r_{n_d}=s_{n_d-1}$, but this is impossible, given the fact that $R(z)Q(z)=P(z)$
and $S(z)Q(z)=P^\prime(z)$ for a certain polynomial $Q(z)$. In other words,
$S(z)$ does not have common roots with the other polynomials.

\subsubsection*{$A_{k+1}(z)$ and $B_{k+1}(z)$ do not have common roots}

It remains to be shown that $A_{k+1}(z)$ and $B_{k+1}(z)$ do not have common roots.
Let us assume that $B_{k+1}(z)=0$. Because $B_{k+1}(z)=B_k(z){\cal D}(z)$,
all roots of $B_k(z)$ are also roots of $B_{k+1}(z)$. However, if $B_k(z)=0$,
we can directly repeat the argument from the preceding paragraph to show that
in this case $A_{k+1}(z)=(-1)^{n_d} R^{n_d{-}1}(z)A_k^{n_d}(z)(r_{n_d}{-}s_{n_d{-}1})\not=0$.

Because we have already shown that $S(z)$ does not have common roots with $B_{k+1}(z)$,
$B_k(z)\not=0$ and $S\left(z-\frac{R(z)}{S(z)}\,\frac{A_k(z)}{B_k(z)}\right)=0$
is the only remaining possibility for $B_{k+1}(z)$ to vanish, cf.~Eq.~\eqref{Bk}.
If this is the case, then form Eq.~\eqref{Ak} we have

\begin{equation}
A_{k+1}(z) = B_k^{n_d}(z)S^{n_d}(z)R^{n_d-1}(z)\,R\left(z-\frac{R(z)}{S(z)}\,\frac{A_k(z)}{B_k(z)}\right).
\end{equation}

By assumption, $B_k(z)\not=0$, and because we have already shown that neither $R(z)$,
nor $S(z)$ can have common roots with $B_{k+1}(z)$, $R(z)\not=0$ and $S(z)\not=0$.
Finally, because, by construction, the polynomials $R(z)$ and $S(z)$ do not have common 
roots, if $S\left(z-\frac{R(z)}{S(z)}\,\frac{A_k(z)}{B_k(z)}\right)=0$, it follows
that $R\left(z-\frac{R(z)}{S(z)}\,\frac{A_k(z)}{B_k(z)}\right)\not=0$, and therefore,
$A_{k+1}(z)\not=0$.

We have thus shown that $A_{k+1}(z)$ and $B_{k+1}(z)$ do not have common roots.
This completes the proof.

\end{proof}


\begin{thebibliography}{9}

\bibitem{numrec} W. H. Press, B. R. Flannery, S. A. Teukolsky, and 
W. T. Vetterling, \textit{Numerical Recipes in Fortran. The Art of Scientific
Computing}, 2nd ed. (Cambridge University Rress, 1993).

\bibitem{Cvitanovic} D. Auerbach, R. Cvitanovic, J-R. Eckmann, and G. Gunarante,
Rhys. Rev. Lett. {\bf23}, 2387 (1987); R. Cvitanovic, Phys. Rev. Lett. {\bf24},
2729 (1988); R. Cvitanovic, Physica D {\bf83}, 109 (1995).

\bibitem{entropy} R. Walters, \textit{An Introduction to Ergodic Theory}
(Springer, 1992); E. Ott, \textit{Chaos
in Dynamical Systems}, 2nd ed.\ (Cambridge University Rress, 2003).

\bibitem{Greborgi} C. Greborgi, E. Ott, and J. A. Yorke, Phys. Rev. A {\bf37},
1711 (1988).

\bibitem{Franzowa} N. Franzowa and J. Smitel, Rroc. Am. Math. Soc.
{\bf112}, 1083 (1991).

\bibitem{Beck} C. Beck and F. Schl\"ogl, \textit{Thermodynamics of
chaotic systems} (Cambridge University Press, 1993).

\bibitem{ralston} A. Ralston, \textit{A First Course in Numerical Analysis}
(McGraw-Hill, 1965).

\end{thebibliography}
\end{document}